\documentclass[]{spie}  

\newcommand{\degc}{$\degree$C}

\usepackage{amsmath,amsfonts,amssymb, gensymb}
\usepackage{multirow}
\usepackage{array}
\usepackage{graphicx}
\usepackage{wrapfig}
\usepackage[normalem]{ulem}
\usepackage[colorlinks=true, allcolors=blue]{hyperref}

\graphicspath{figures}

\title{Fabrication status and expected performance of the inner-core X-ray optic for BabyIAXO}

\author[a]{Jooyun Woo}
\author[a]{Yue Yu}
\author[a]{Ipek Altunyurt}
\author[a]{Todd Decker}
\author[b,c]{Desiree Della Monica Ferreira}
\author[b]{Peter Lindquist Henriksen}
\author[a]{Eftychia Kotsiou}
\author[b,c]{Sonny Massahi}
\author[a]{Kerstin Perez}
\author[a]{Anacorina Romero}
\author[b]{Jaime Ruz}
\author[a,d]{Vyshnavi Sabbi}
\author[a]{Marcela Stern}
\author[e]{Julia Katharina Vogel}
\affil[a]{Columbia University, 550 W 120th St, New York, NY 10027, USA}
\affil[b]{Technical University of Denmark, 2800 Kgs. Lyngby, Denmark}
\affil[c]{CHEXS, 2800 Kgs. Lyngby, Denmark}
\affil[d]{University of North Carolina at Chapel Hill, Chapel Hill, NC 27599, USA}
\affil[e]{TU Dortmund University, 44227 Dortmund, Germany}

\authorinfo{Send correspondence to Jooyun Woo: jw3855@columbia.edu}

\begin{document} 
\maketitle

\begin{abstract}
BabyIAXO, a pathfinder for the International Axion Observatory (IAXO), is designed to demonstrate all key technologies at scale while achieving an improvement in sensitivity over the recent CERN Axion Solar Telescope (CAST) experiment by approximately a factor of five. Such improvement is enabled by the X-ray optics, which allow for maintaining a high signal-to-noise ratio at the detector despite a cross-sectional area of the magnetic bore being over $250$ times larger than that of CAST. The optic employs a hybrid design consisting of co-aligned inner core and outer corona optics that share a common optical axis and vacuum vessel but differ in focal length and manufacturing approach. Both are segmented glass optics, with the inner core fabricated from thermally slumped borosilicate glass and the outer corona from cold-slumped Corning Willow glass. To fabricate the inner-core optic, leveraging techniques developed for NuSTAR and HEFT optics, we reoptimized and streamlined the thermal-forming procedure. The quality of free-standing glass substrates was characterized by laser metrology, X-ray reflectometry, and atomic force microscopy. We developed a cutting technique that produces smooth edges at the micron scale. We used flat stacks of glass-epoxy-graphite layers to evaluate the performance of the epoxy bondline. The optic is expected to achieve an on-axis point spread function (PSF) with a half-power diameter (HPD) of $<90''$, enhancing the signal-to-noise ratio by more than 55 times.
\end{abstract}


\section{INTRODUCTION}
\label{sec:intro}  

BabyIAXO \cite{AA21:babyIAXO_cdr}, a pathfinder for the International Axion Observatory (IAXO \cite{II2011:iaxo_instrument,EA19:iaxo_physics}), is an axion helioscope experiment designed to detect axions and axion-like particles produced in the solar core (``solar axions'') that are converted into X-ray photons via strong magnetic fields. BabyIAXO will host two magnetic bores, each with a field strength of $\sim2$ T and a diameter of 700\,mm, a cross-sectional area 250 times larger than that of the CERN Axion Solar Telescope (CAST) experiment, which delivered the current leading helioscope sensitivity \cite{MA15:cast, VA17:cast}. Each magnetic bore will be equipped with a focusing X-ray optic to reduce the spot size and improve the signal-to-noise ratio. BabyIAXO will increase the sensitivity to the axion-photon coupling by a factor of 5 compared to CAST, with IAXO improving by over a factor of 20.

A custom optic, a pathfinder for the IAXO optics, is under construction for one of the magnet bores, while the flight-spare optics module for XMM-Newton \cite{FJ2001:xmm} will be used for the other. The BabyIAXO custom optic adopts a hybrid design comprising an inner core optic, being developed at Columbia University, and an outer corona optic, being developed at INAF/OABrera. The two optics are co-aligned along a common optical axis and housed within the same vacuum vessel, while employing different focal lengths and fabrication techniques. Both are segmented glass optics: the inner core optic is constructed from thermally slumped borosilicate glass \cite{JK03:heft_optics, JK04:nustar, JK05:nustar, JK09:nustar, CH10:nustar, WC11:nustar, WZ09:nustar_optics}, whereas the outer corona optic uses cold-slumped Corning Willow glass \cite{MC13:outer_corona, MC16:cold_slumping}. In this proceeding, we describe the fabrication process for and the development status of the custom inner-core optics.


\section{OPTIMIZED DESIGN AND REQUIREMENTS}
\label{sec:design}

The BabyIAXO (and IAXO) optic is designed for the detection of weak axion-induced X-ray signals, leading to design requirements that differ significantly from those of conventional X-ray telescopes. Because axions are expected to originate primarily from the inner $\sim3'$ of the solar core, maximizing effective area is prioritized over achieving high spatial resolution. In addition, the optic can operate on the ground, eliminating launch-related constraints and simplifying material selection and environmental qualification.

At the same time, a large-area ground-based X-ray optic presents several challenges. The optic must 
withstand gravitational loads that vary as the telescope tracks the Sun. The coatings must be optimized for the expected axion spectra, and the materials must satisfy stringent radiopurity requirements because the optic shares a vacuum vessel with the detectors. Furthermore, as a pathfinder for IAXO, which will require eight magnetic bores and matching optics, the design must be scalable and cost-effective.

The inner-core X-ray optic of BabyIAXO has its heritage in the construction method developed for High-Energy Focusing Telescope (HEFT \cite{JK03:heft_optics,JK04:heft,JK06:heft}), Nuclear Spectroscopic Telescope Array (NuSTAR \cite{JK04:nustar, JK05:nustar, JK09:nustar, CH10:nustar, WC11:nustar}), and the prototype optic for CAST \cite{MP26:cast_optics}. The optic adopts a conical approximation of the Wolter-I design, where the primary and secondary mirror shells of the Wolter-I design are approximated as truncated cones. Construction utilizes a simple, but high-performance, technique in which thermally slumped glass segments are individually coated with reflecting materials and precisely mounted by mechanical means into nested shells of reflecting surfaces. 

Table \ref{tab:design} lists the design parameters that specify the inner-core optic. Derivation of this design is detailed in \cite{YY26:design}. Ray-tracing simulations assuming ideal mirror figure yield an on-axis point spread function (PSF) with a half-power diameter (HPD) of $\sim46''$, driven by the conical approximation.
To reduce cost, we leverage the equipment and facilities used for HEFT and NuSTAR, and combine custom mirrors with spare NuSTAR glass and coated flight mirrors. Accordingly, the BabyIAXO inner-core optic adopts the same mirror length as NuSTAR (225 mm), with shell radii chosen to closely match NuSTAR's radial range. Recent reflectivity measurements at SOLEIL synchrotron beamlines, as well as theoretical studies demonstrating the suitability of NuSTAR coatings for BabyIAXO \cite{YY26:design}, have confirmed that the coatings of the NuSTAR mirrors retain excellent performance down to $<1$ keV even after more than a decade of storage. 
The custom substrates are fabricated using the thermal-forming process developed for HEFT \cite{MJ03:thermal_forming}. Because the available furnaces cannot accommodate the full 225\,mm mirror length, individual segments are slumped to approximately half that length (112\,mm) and subsequently paired to form a complete primary or secondary mirror.

\begin{table}[t!]
    \centering
    \begin{tabular}{|m{0.4\textwidth}|m{0.45\textwidth}|}
        \hline
        \textbf{Parameter} & \textbf{Value} \\
        \hline
        Focal length & 5.6 m \\
        \hline
        Number of concentric shells & 103 \\
        \hline
        Axial length of segment & 112\,mm (custom) or 225\,mm (spare NuSTAR) \\
        \hline
        Axial length of primary or secondary mirror & 225 mm \\
        \hline
        Diameter range & 108-382 mm \\
        \hline
        Number of shells: 60$\degree$ segments & Inner 54 shells \\
        \hline
        Number of shells: 30$\degree$ segments & Outer 49 shells \\
        \hline
        Substrate type & Borosilicate glass (D 263 T eco) \\
        \hline
        Substrate thickness & 0.21 mm \\
        \hline
        Total number of mirrors & 1,824 \\
        \hline
        Coating & Five recipes of Pt/C bilayer \\
        \hline
    \end{tabular}
    \caption{Key design parameters for the inner-core X-ray optic for BabyIAXO \cite{YY26:design}.}
    \label{tab:design}
\end{table}

\section{FABRICATION STATUS}

\begin{figure} [t!]
\begin{center}
\begin{tabular}{cc}
\includegraphics[width=0.95\textwidth]{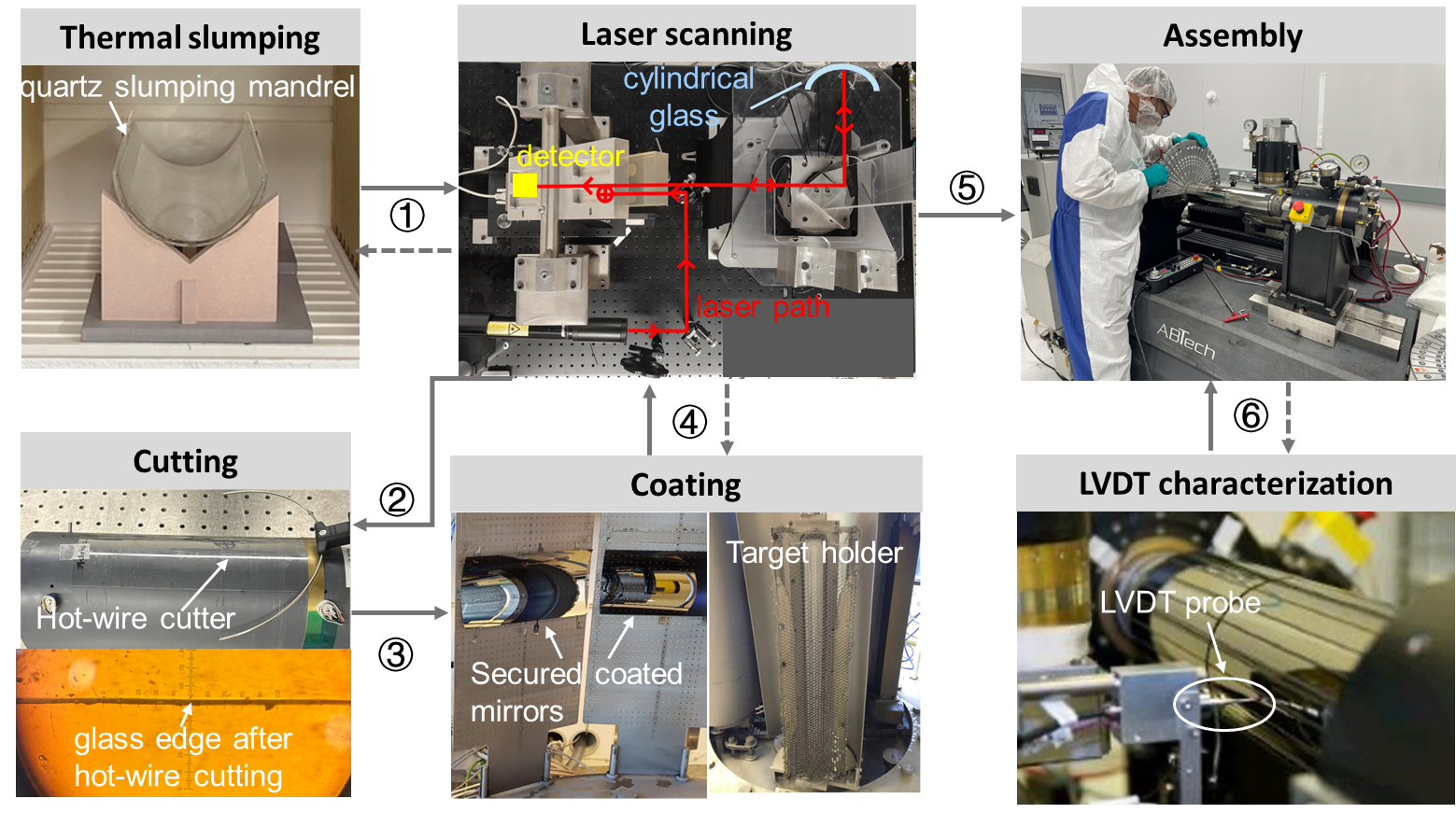} 
\end{tabular}
\end{center}
\caption{\label{fig:workflow}
BabyIAXO inner-core optics fabrication process \cite{YY26:design}.}
\end{figure}

The inner-core optics fabrication process is shown in Figure \ref{fig:workflow}. Flat glass sheets are thermally slumped into cylindrical mandrels in a high-temperature furnace and subsequently cut into trapezoidal segments using a combination of diamond-scribing and hot-wire techniques. Free-standing glass figure is confirmed using laser scanner metrology. The resulting substrates are then coated with high-reflectivity films. In the following assembly stage, glass figure errors are mitigated using the error-correcting monolithic assembly and alignment (EMAAL) technique \cite{CH97:emaal,JK03:emaal,JK09:emaal}, which replaces conventional optical alignment with a mechanical process. Using a dedicated assembly machine, the glass segments are epoxied to precisely machined graphite spacers, forcing the initially cylindrical substrates into the required conical geometry and correcting radial mismatches and small twists in the glass. The mounted glass figure is measured by scanning a Linear Variable Differential Transformer (LVDT) over the back surface of the mirror. Each layer of spacers is machined
with respect to the optic axis and not the last layer of glass, eliminating any stack-up error. 
Glass production, characterization, and optic assembly are carried out at Columbia University. Slumped and cut glass segments will be coated with a Pt/C bilayer at Technical University of Denmark.


\subsection{Substrate production}
\subsubsection{Thermal forming}
\label{subsubsec:slumping}


The mirror substrate production is based on a glass slumping technique developed for HEFT \cite{JK06:heft, MJ03:thermal_forming}. In this technique, commercially available thin borosilicate glass sheets are slumped, due to gravity and changes in viscosity, into convex quartz mandrels until shortly before the glass comes into contact with the mandrel. The resulting glass figure largely follows the overall cylindrical shape of the mandrel, while preserving the excellent microroughness of the glass sheet (RMS $\leq1$ nm) without inheriting the mandrel's surface figure error. 

We use Vulcan 3-1750 programmable furnaces with heating elements on the left and right walls. 
The spatial and temporal temperature profiles of each furnace were measured for quality control and cross-calibration. A 2.5 mm-thick convex quartz mandrel is placed on ceramic stands inside the furnace. Mandrel diameters, on average, are within 1.35 mm of the design values and range from 108 mm to 384 mm in 4-mm or 5-mm intervals. A piece of silica fabric is placed on top of the mandrel surface as a release layer. 
A flat glass sheet (Schott D263 T eco) of dimensions 361\,mm $\times$ 439\,mm $\times$ 0.21\,mm is cut via diamond scribe into segments 200\,mm long (along the draw direction and the optical axis) and 5\,mm wider than the mandrel diameter (to account for the mandrel thickness) and placed along the edges of the mandrel. 

The furnace is programmed with three stages: ramp-up, thermalization, and forming. During the ramp-up and thermalization, the temperature is increased to the strain point of the glass (529\degc) at different rates (15\degc\ and 9\degc\ per minute, respectively) and held for 10 minutes. The temperature is then increased at a slower rate 
\begin{wrapfigure}{r}{0.5\textwidth}
\includegraphics[width=0.49\textwidth]{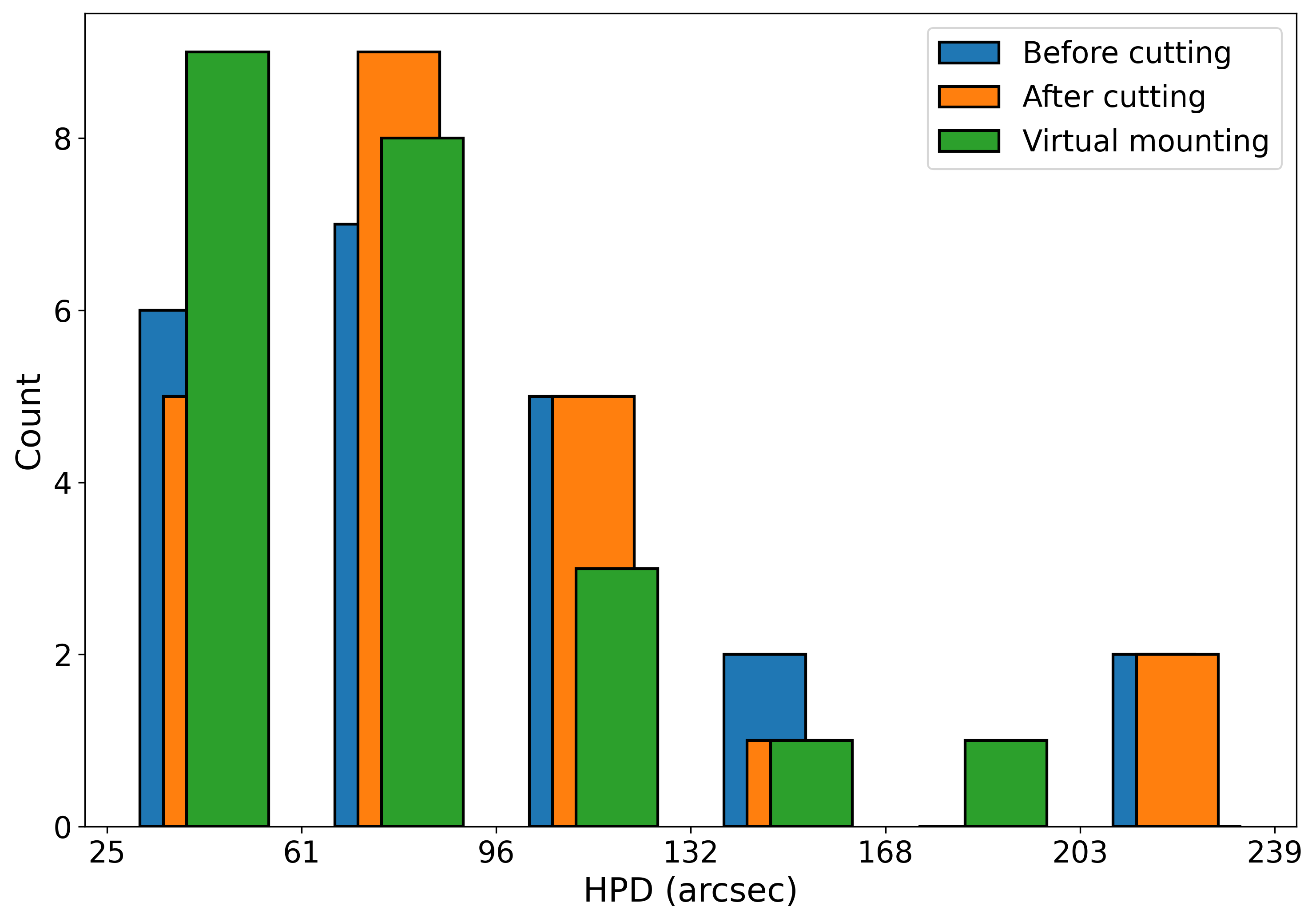}
\caption{\label{fig:scanning}
Histogram of HPDs before and after cutting, and after virtual mounting. The distribution includes glass segments produced before the optimization of soaking temperature and time.}
\end{wrapfigure} 
(5\degc\ per minute) to the ``soaking temperature'' and held for the ``soaking time'', during which glass forming occurs. The soaking temperature, between the annealing (557\degc) and softening (736\degc) points, and the soaking time are determined for each mandrel diameter and furnace, informed by our measurements of furnace temperature profiles, the viscodynamic model \cite{MJ03:thermal_forming}, and the resulting glass figure characterized by laser metrology.
We achieve a 70\% yield for the target mirror figure error of $<76''$ (HPD) by optimizing the soaking temperature and time. 


\subsubsection{Glass segment cutting}
\label{subsubsec:cutting}


Glass segments must then be cut to the required geometry for mounting. In addition, glass segments exhibit large figure errors along the edges caused by contact with the mandrel (in the axial direction) and a high gradient in furnace temperature (in the azimuthal direction). Therefore, we cut out the central part of the slumped segments in the required trapezoidal shape.

Glass segments are cut with a combination of diamond scribe and hot-wire techniques. We create templates for each layer with 0.1-mm precision. A glass segment is secured on top of template, which is attached to a cutting mandrel (PVC pipes of varying diameters), and the corners of the template are marked on the glass. First, the axial edges are cut with a diamond-tipped scriber along a guiding metal rod. Next, the top and bottom edges are scored with a diamond-tipped scriber along the axial cutting stencil, and then cut with a handheld hot-wire cutter. The wire (32 AWG Nichrome 80) and the operating voltage (9V) were chosen so that the wire temperature is below the strain point yet sufficient to induce thermal shock. Both the axial and azimuthal cutting result in smooth edges to the micron scale. The cutting procedure also achieves a yield of 75\%, defined as the fraction of segments successfully cut without breakage. 

As a ground mission, the most significant vibration our X-ray optics will experience during fabrication and operation is from pre-coating cleaning in an ultrasonic cleaner; this is unlike space missions, where optics need to withstand much stronger vibration loads during launch. We have verified the robustness of the cut edges to mechanical stresses induced by the ultrasonic cleaning procedure. 


\subsubsection{Laser metrology and virtual mounting simulation}
\label{subsubsec:laser_scanning}



The resulting glass figure error is predominantly determined by surface errors with a spatial frequency of  $\mu m$ to cm scale along the optical axis. These errors are measured using the laser metrology system developed for NuSTAR \cite{HA11:nustar_thesis, MJ98:llaser_scanner}.

When scanning pre-coating segments, the back surface is painted with a 2:1 mixture of sugar and black tempera paint to suppress back-surface reflections. The sugar closely matches the refractive index of the glass, while the paint absorbs transmitted light. \cite{MJ00:sugar}.  
The segment is then scanned to produce a height profile of the optical surface at the micron scale and to calculate the half-power diameter (HPD). The height profile and HPD provide guidance on optimizing soaking temperature and time during slumping, thereby improving the yield.

\begin{wrapfigure}{r}{0.6\textwidth}
\includegraphics[width=0.6\textwidth]{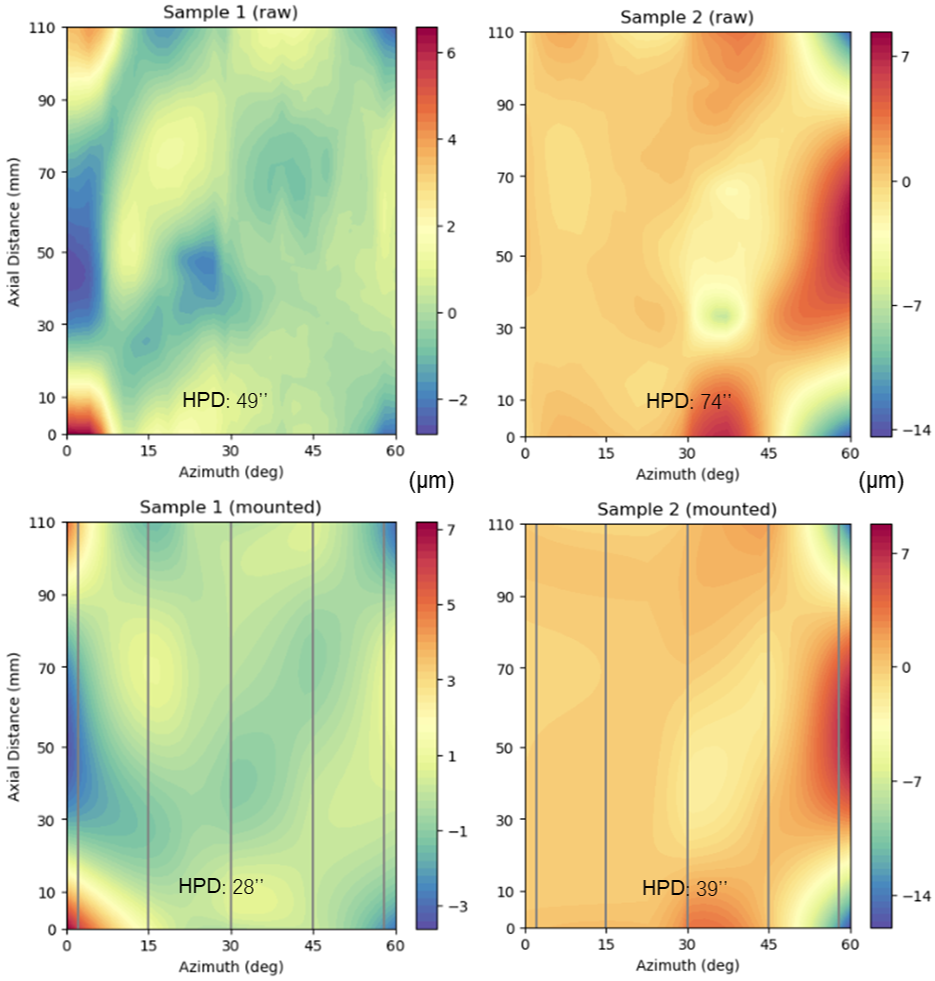}
\caption{\label{fig:vm}
2D surface profiles of two slumped segments measured with the laser scanner (top) and after virtual mounting (bottom). The grey vertical lines in the bottom panel indicate the location of the spacers \cite{YY26:design}.}
\end{wrapfigure}

The mirror surface figure measured with the laser scanner is used to estimate the mirror's optical performance after mounting. The HPD is expected to improve after mounting, as conic–cylindrical mismatch is corrected and the overall axial bow is largely removed, redistributing any residual error into shorter-wavelength figure errors. We simulate this effect using our virtual mounting code by assuming that a mirror's surface figure errors on top of the spacers are perfectly corrected, and that the remaining figure errors are redistributed across the surface between the spacers. The upper panel of Figure \ref{fig:vm} presents the figure errors of two representative glass segments before mounting, while the lower panel shows the corresponding figure errors after virtual mounting. Sample 1 highlights the quality achievable through the slumping and cutting processes, with an HPD of $49''$, 
which is further improved to $28''$ after virtual mounting. Sample 2 demonstrates the larger corrections that can be achieved when the segment with a higher HPD of $74''$ is constrained by the spacers, reducing the HPD to $39''$. A systematic improvement in HPD post-mounting is observed as shown in Figure \ref{fig:scanning}. The segments with estimated post-mounting HPD $<76''$ are selected for coating. 

\subsubsection{Microroughness measurement}
\label{subsubsec:roughness}

Microroughness with spatial wavelengths ranging from a few nm to $\mu$m is a dominant factor in X-ray scattering and in the degradation of performance in multilayer coatings. The excellent microroughness of the flat glass sheet ($<1$ nm RMS) may not be preserved during the optics fabrication processes, such as slumping, cleaning, and coating. We measured the microroughness of glass samples at two stages, after slumping and after cleaning, using two independent methods, atomic force microscopy (AFM) and X-ray reflectometry (XRR). The microroughness measured at both stages with both methods yielded consistent values with those of the flat glass sheet. After coating, the microroughness is estimated to contribute approximately $5''$ to the optic HPD through X-ray scattering. \newline

\subsection{Mirror Mounting}
\label{subsec:epoxy}

We use the same assembly machine, technique, and material (Trabond F131 epoxy, graphite spacer, titanium innermost mandrel) as NuSTAR. In the EMAAL process, the assembly machine rotates the optic at 100 rpm and machines the spacers into the required conical profile relative to the optical axis, rather than previously assembled mirror layers, thereby eliminating stack-up errors. This process strictly controls the relevant sources of HPD degradation, achieving spacer-machining accuracy $<8''$ and mandrel repeatability $<5''$ \cite{CH97:emaal,JK03:emaal,JK09:emaal}. In contrast, epoxy application is performed manually, and the resulting bondline uniformity and mechanical properties depend sensitively on both workmanship and curing conditions.

Epoxy affects the optical and structural performance in three ways. First, the uniformity of the epoxy bondline directly contributes to the surface figure error of the mirror in contact. Second, the mechanical strength of the epoxy bond determines the durability of the optics structure under stresses introduced during operation. Lastly, the outgassing of the epoxy increases X-ray scattering. We present the results of epoxy bondline uniformity and mechanical strength testing to validate the process, material, and workmanship, and refer to previous work that confirmed requisite low-outgassing performance\cite{HA09:epoxy}. 

\begin{figure} [t]
\begin{center}
\begin{tabular}{cc}
\includegraphics[width=0.95\textwidth]{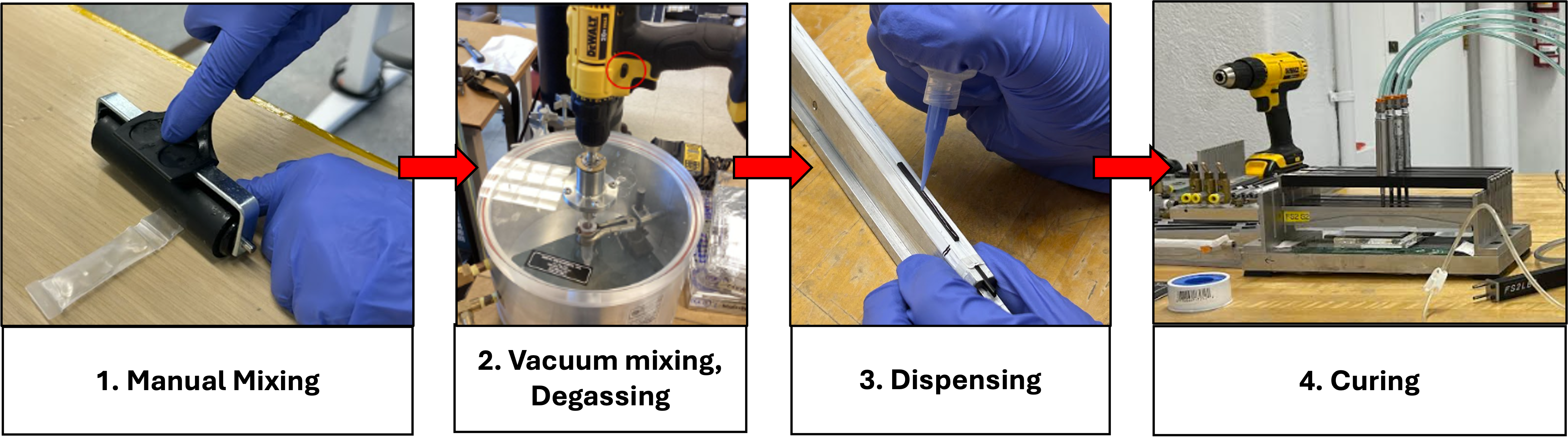} 
\end{tabular}
\end{center}
\caption{\label{fig:epoxy}
Flat-stack preparation process for epoxy performance testing \cite{YY26:design}.}
\end{figure}

\begin{figure} [b!]
\begin{center}
\begin{tabular}{cc}
\includegraphics[width=0.95\textwidth]{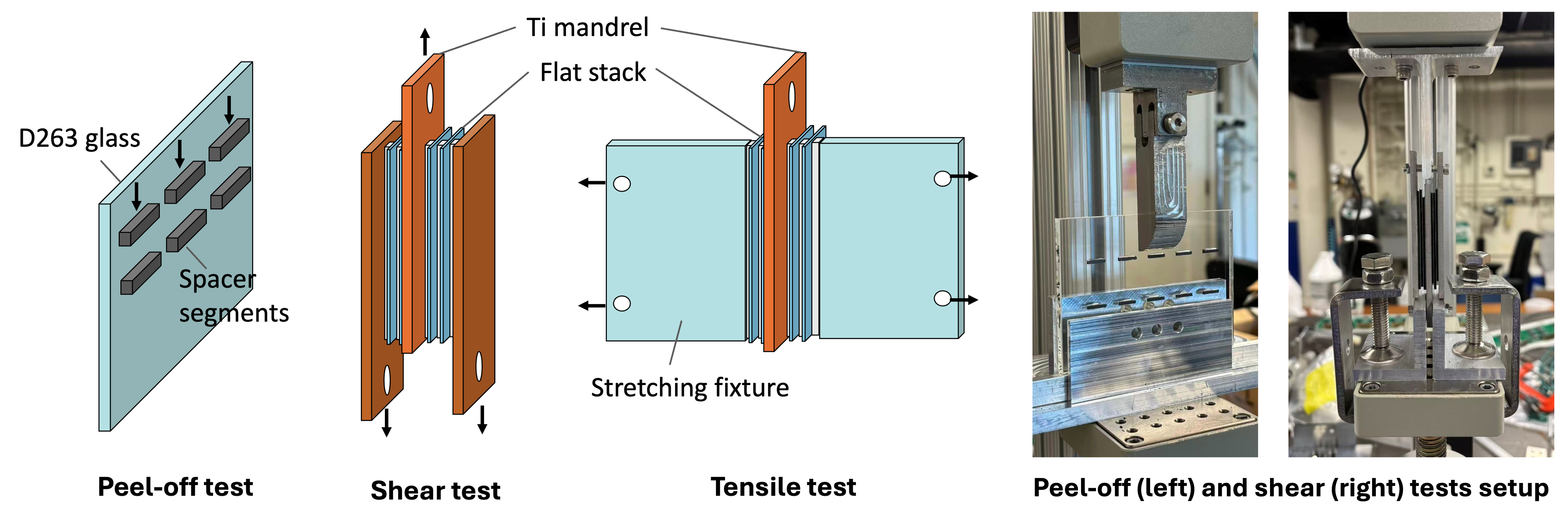} 
\end{tabular}
\end{center}
\caption{\label{fig:strength}
Setups for peel, shear, and tensile tests. The black arrows in the first three schematic diagrams show the direction of the applied force. The last two pictures show the flat stacks mounted on the testing aparatus.}
\end{figure}

Epoxy bondline and strength studies are carried out on flat stacks, hand-sized testing samples fabricated using the same materials, assembly process, and linear density of epoxy as the final optics (Figure \ref{fig:epoxy} and \ref{fig:strength}). 
We validate the optimal epoxy mass density (determined by dispensing speed and pressure) and curing environment (temperature, humidity, pressure, and time) through peel tests.
Under the given epoxy formulation, the bondline uniformity and structural strength are directly determined by the epoxy mass density, pot life, and curing pressure. In addition, the curing temperature, duration, and ambient relative humidity also significantly affect the curing speed and adhesion quality. 
Flat stacks are produced under varying epoxy and environmental parameters, and then spacers are peeled off to measure the failure load and mode. The optimal parameters that yield the highest failure load and predominantly cohesive failures within the epoxy (rather than adhesive failures at the epoxy-sample interface) are adopted. Accordingly, flat stacks were prepared in a temperature-controlled (room temperature) and humidity-controlled ($<30$\% relative humidity) dry room, with an epoxy curing pressure of $\sim15$ psi and a curing time that varied with temperature (minimum 3 days).
Glass and spacer samples are thoroughly cleaned and dried overnight before use.


The uniformity of the epoxy bondline is measured using a Linear Variable Differential Transformer (LVDT), a high-resolution ($\sim$15 nm), low-force ($\sim$1g) mechanical probe. We obtained a consistent uniformity of $< 0.2\ \mu$m for a bondline thickness of $<2\ \mu$m, which corresponds to an HPD degradation of $<3''$ \cite{HA09:epoxy}. Combined with other error sources, such as spacer figure errors and cylindrical-to-conic mirror deformation, the overall contribution of the mounting process to the HPD is estimated to be $\sim15''$.

The mechanical strength of the epoxy bondline is measured through shear tests and tensile tests (Figure \ref{fig:strength}). The flat stacks for those tests are constructed to mimic the actual stress-loading conditions of the telescope's inner layers. The planar titanium pieces are prepared using the same procedures as the innermost mandrel for the full optics. Graphite-epoxy-glass composite layers are constructed on both sides of a titanium piece. For shear testing, another titanium piece is epoxied on top of the composite layers.
Our test results yield a mean failure load of 423 lbs and a maximum of 620 lbs for the shear tests, and 100-200 lbs for the tensile tests. 



\section{SUMMARY}
\label{sec:summary}

We have optimized the fabrication process of the inner-core X-ray optic for BabyIAXO to achieve the target mirror figure error of $<76''$ (HPD), which, combined with the independent figure errors introduced by the design \cite{YY26:design} and other fabrication steps (Table \ref{tab:error}), yields the on-axis PSF of HPD $<90''$ and the signal-to-noise ratio enhanced by more than 55 times \cite{YY26:design}. A 10-layer prototype is scheduled for assembly in mid-2026 and calibration at PANTER \cite{JK11:nustar_calibration, NB11:ramcaf, NB12:nustar_calibration} later that year. If its measured performance matches expectations, the prototype will be incorporated as the inner 10-layer module of the full telescope. Construction of the full inner optic is anticipated to be completed by late-2027. 

\begin{table}[h!]
\centering
\begin{tabular}{l|r}
\hline
Error source & HPD contribution ($''$) \\
\hline
Conic approximation & 46 \\
Mirror surface figure error & $<76$ \\
Mounting figure error & $\sim15$ \\
Scatter from microroughness & $\sim5$ \\
\hline
Total PSF HPD & $<90$ \\
\hline
\end{tabular}
\caption{\label{tab:error}
Error budget for the inner-core optic.}
\end{table}

\acknowledgments 
 
We thank Chuck Hailey for helpful discussions. This research was supported by the National Science Foundation (NSF) PHY-2309980 and NSF Research Experiences for Undergraduates (REU) Program under Grant NSF PHY-2349438. 

\bibliography{references} 
\bibliographystyle{spiebib} 

\end{document}